\documentclass[11pt]{article}
\usepackage{hyperref}
\pdfoutput=1

\begin{document}
\title{Light Enabled Digital Microfluidics:A Technology Leading to a Programmable Lab on a Chip}
\author{Han-Sheng Chuang, Aloke Kumar, Steven T. Wereley \\
\\\vspace{6pt} School of Mechanical Engineering,
\\ Purdue University, West Lafayette, IN 47907, USA}
\maketitle
\begin{abstract}
This fluid dynamics video showcases how optically induced electrowetting can be used to manipulate liquid droplets in open space and particulate phases inside the droplet. A photoconductive layer is added to a conventional electrowetting-on-dielectric (EWOD) structure to generate light enabled virtual electrodes, hence resulting in an eletrowetting action. Coplanar electrodes deployed alternately on a substrate enable open droplet manipulations differentiating from a sandwiched configuration. An integration with an optoelectric method shows dynamic and rapid particle handling by strong micro fluidic vortices in conjunction with other electrokinetic forces inside a droplet. The droplet manipulations are realized with visible illumination and powered at 150 volts peat-to-peak with a low frequency (100 Hz-800 Hz). The particle concentration is achieved on the surface of the same chip but illuminated with a near-infrared (1064 nm) light source and biased with a high frequency (24 kHz) AC signal. 
\end{abstract}
\section{Introduction}

The sample video is
\href{http://ecommons.library.cornell.edu/bitstream/1813/14103/3/O-OEW_movie_MPEGI.mpg}{Video1} \\
\\

We present an open optoelectrowetting (O-OEW) technique for droplet manipulations. The proposed O-OEW features dynamic droplet maneuverability and great extensibility due to light-induced virtual electrodes and an open configuration. 
The device comprises coplanar interdigitated electrodes, a photoconductor, and an insulator on a substrate. 
The mechanism behind the O-OEW is dependent on the impedance switching between the photoconductor and the 
insulator. The photoconductor works as a binary gate for the equivalent circuit. When illuminated the impedance of 
the photoconductor decreases, prompting an electrowetting effect due to a high voltage drop in the insulator. 
Without illumination the impedance of the photoconductor increases, shifting the voltage drop back to the 
photoconductor layer and turning off the electrowetting effect. Illumination induces a localized hydrophilic region 
on an overall hydrophobic surface, causing an imbalance of surface tension forces and the subsequent liquid droplet 
movement. By selectively illuminating the platform surface, basic droplet operations, such as 
translation, merging, and simultaneous multi-droplet control, are implemented. Immersing the liquid droplets in oil enhances 
the movements and prevents serious evaporation. For more high-end applications, an addressable light source or other tecniques can be integrated to broaden its applications. The integration will enable the realization of a programmable lab-on-a-chip system.
\\
In this video, water droplets of around 20 $\mu$L are manipulated in an silicon oil environment with either a broadband illumination mercury lamp (100 W) or a He-Ne laser head of no more tham 20mW. The applied electric potential is 150 volts peak-to-peak and the AC frequency is optimized between 100 Hz and 800 Hz according to our previous study [1]. The first section shows the basic optoelectrowetting of a dropet (not immersed in oil) in response to the illumination from a laser pointer. The second section demonstrates some droplet manipulations, including translation, merging, and simultaneous multi-droplet handling. Especially, the low-viscous silicone oil can be split and merge guided by a water droplet because the affinity between silicone oil and water molecules is stronger than the cohesion of the oil molecules. In the last section, O-OEW combining with an optoelectric method [2] is exhibited for showing a more practical programmable lab-on-a-chip applications. A water droplet containing carboxyl modified particles of 1.0 $\mu$m is moved toward an observation site (not shown in the video) where particle concentration will be performed. Then the particles inside the droplet are rapidly and dynamically concentrated with optoelectrothermal vortices in conjunction with other electrokinetic forces. In contrast to O-OEW, a low electric potential (28 volts root-mean-square) and a high AC frequency (24 kHz) are applied to the concentration.\\

\section{References}
\begin{enumerate}
\item
H.S. Chuang, A. Kumar, S.T. Wereley (2008) Open Optoelectrowetting Droplet Actuation, Appl. Phys. Lett. 93: 064104.   
\item
S.J. Williams, A. Kumar, S.T. Wereley (2008) Electrokinetic patterning of colloidal particles with optical landscapes, Lab Chip 8: 1879-1882.  
\end{enumerate}
\end{document}